\documentclass[sigconf]{acmart}
\copyrightyear{2025}
\acmYear{2025}
\setcopyright{cc}
\setcctype{by}
\acmConference[UMAP '25]{33rd ACM Conference on User Modeling, Adaptation and Personalization}{June 16--19, 2025}{New York City, NY, USA}
\acmBooktitle{33rd ACM Conference on User Modeling, Adaptation and Personalization (UMAP '25), June 16--19, 2025, New York City, NY, USA}
\acmDOI{10.1145/3699682.3728330}
\acmISBN{979-8-4007-1313-2/2025/06}

\def\plainauthor{Kayhan Latifzadeh, Luis A. Leiva, Klen Čopič Pucihar, Matjaž Kljun, Iztok Devetak, Lili Steblovnik}
\def\plaintitle{Assessing Medical Training Skills via Eye and Head Movements}
\def\plainkeywords{eye movements, head movements, simulation training}

\usepackage[utf8]{inputenc}
\usepackage{wasysym}
\usepackage{bm}
\usepackage{url}
\usepackage{colortbl}
\usepackage{booktabs}
\usepackage{multirow}
\usepackage{makecell}
\usepackage{tablefootnote}
\usepackage{enumitem}
\usepackage{soul} 
\usepackage{xcolor}

\usepackage{graphicx}
\usepackage{tabularx}
\usepackage{multirow}
\usepackage{subcaption}

\graphicspath{={./figs/}}

\sethlcolor{red}

\hypersetup{%
  pdftitle={\plaintitle},
  pdfauthor={\plainauthor}
  pdfkeywords={\plainkeywords},
  bookmarksnumbered,
  pdfstartview={FitH},
  colorlinks,
  allcolors=black,
  breaklinks,
}

\begin{document}

\title{Assessing Medical Training Skills via Eye and Head Movements}

\author{Kayhan Latifzadeh}
\authornote{Authors contributed equally to this research.}
\orcid{0000-0001-6172-0560}
\affiliation{\institution{University of Luxembourg}
\city{Luxembourg}
\country{Luxembourg}}
\email{kayhan.latifzadeh@uni.lu}

\author{Luis A. Leiva}
\authornotemark[1]
\orcid{0000-0002-5011-1847}
\affiliation{\institution{University of Luxembourg}
\city{Luxembourg}
\country{Luxembourg}}
\email{luis.leiva@uni.lu}

\author{Klen \v{C}opi\v{c} Pucihar}
\authornotemark[1]
\orcid{0000-0002-7784-1356}
\affiliation{\institution{University of Primorska}
\city{Koper}
\country{Slovenia}}
\affiliation{\institution{Stellenbosch University}
\city{Stellenbosch}
\country{South Africa}}
\email{klen.copic@famnit.upr.si}

\author{Matja\v{z} Kljun}
\orcid{0000-0002-6988-3046}
\affiliation{\institution{University of Primorska}
\city{Koper}
\country{Slovenia}}
\affiliation{\institution{Stellenbosch University}
\city{Stellenbosch}
\country{South Africa}}
\email{matjaz.kljun@upr.si}

\author{Iztok Devetak}
\orcid{0000-0003-4719-8424}
\affiliation{\institution{University of Ljubljana}
\city{Ljubljana}
\country{Slovenia}}
\email{iztok.devetak@pef.uni-lj.si}

\author{Lili Steblovnik}
\authornotemark[1]
\orcid{0000-0001-8868-7721}
\affiliation{\institution{University Medical Centre Ljubljana}
\city{Ljubljana}
\country{Slovenia}}
\email{lili.steblovnik@mf.uni-lj.si}

\renewcommand{\shortauthors}{K. Latifzadeh et al.}

\begin{abstract}
We examined eye and head movements to gain insights into skill development in clinical settings.
A total of 24 practitioners participated in simulated baby delivery training sessions. 
We calculated key metrics, including pupillary response rate, fixation duration, or angular velocity.
Our findings indicate that eye and head tracking can
effectively differentiate between trained and untrained practitioners, particularly during labor tasks. 
For example, head-related features achieved an F1 score of 0.85 and AUC of 0.86, 
whereas pupil-related features achieved F1 score of 0.77 and AUC of 0.85.
The results lay the groundwork for computational models 
that support implicit skill assessment and training in clinical settings 
by using commodity eye-tracking glasses
as a complementary device to more traditional evaluation methods
such as subjective scores.
\end{abstract}


\maketitle

\section{Introduction}
\label{sec:introduction}

Simulation based training plays a crucial role in preparing specialized medical professionals by providing a safe environment for skill development and hands-on experience~\cite{ccalim2020effect, mannella2018simulation, cai2022towards}. This includes both technical and non-technical skills, such as communication and teamwork~\cite{brandao2018importance, saleem2023healthcare}. Practicing these skills in a controlled setting reinforces muscle memory and builds confidence, which can lead to improved performance in real-life situations. 

In addition, simulated medical environments enable instructors to deliver immediate feedback, supporting reflective learning. This process encourages practitioners to critically examine their experiences, thoughts, and responses, allowing them to gain deeper insights and improve future performance. This method facilitates more effective learning compared to or in combination with traditional classroom-based learning. However, it requires continuous observation, which can be time-intensive and prone to human biases~\cite{buhler2024task, shah2001simulation, das2024shifting}.

Nowadays, integrating biosignals into simulation training offers several benefits that can enhance learning outcomes and overall training effectiveness~\cite{faust2018deep, ali2023medical}. These signals offer an implicit and objective means of skill assessment, providing faster and less biased evaluation. Among the various biosignals, eye and head movement tracking are commonly used, as they can be collected unobtrusively using lightweight wearable devices such as eye and head tracking glasses~\cite{meyer2022u, onkhar2024evaluating, hafner2023eye, huang2024assessing, cognolato2018head}. 
Despite numerous studies exploring the use of eye tracking and head movement in simulation training, the impact of different features from eye (e.g., pupil size, fixation, and saccades) and head movements (velocity and rotation) is still poorly understood. Since these factors may influence different tasks in varying ways, further asessment is necessary~\cite{valtakari2021eye}.

This study investigates the role of eye and head movements in assessing skill acquisition during simulation based training for medical professionals in the context of breech delivery--a childbirth scenario that has not yet been explored. By addressing this gap, we contribute insights into how eye and head tracking technologies can enhance both the assessment and training of healthcare professionals, ultimately leading to improved patient care during childbirth. Our key contributions include:

\begin{itemize}[leftmargin=13pt, topsep=4pt]
    \item Exploration of the potential of eye and head movements as indicators of skill acquisition in medical training.
    \item Identification and analysis of key metrics from eye and head movements for this purpose.
    \item Introduction of an eye/head movements dataset~\footnote{\url{https://zenodo.org/records/15163456}} of 48 breech delivery procedures,
        worth of over 8\,h of movement data,
        featuring annotated time segments and post-session skill scores. We also release python scripts, and preprocessed data~\footnote{\url{https://github.com/kayhan-latifzadeh/LaborTrack}}. 
\end{itemize}

\section{Background and related work}
\label{sec:relatedwork}

Eye-tracking glasses have become a valuable tool in medical research~\cite{ibragimov2024use, bapna2023eye, zhu2024wearable, ferreira2024advancing}, providing insights into the performance of healthcare professionals during task execution. 
These devices enable researchers to evaluate key metrics such as fixations, saccades, or pupil size, 
which have been, for example, linked to decision-making~\cite{al2017eye} and attention~\cite{hofmaenner2021use}.
These metrics can be used for modeling skill development and adapting content in medical training. For example, studies have demonstrated that healthcare professionals can use eye-tracking metrics to identify areas for improvement and optimize simulation-based learning procedures for both novice and expert practitioners~\cite{skaramagkas2021review, souchet2022measuring, hoffman2016visual}. 
In this section we review related work on eye and head tracking in clinical settings,
and outline the research hypotheses we aim to validate through a user study.

\subsection{Eye tracking}

Traditionally, eye tracking research has used Areas of Interest (AOI) analysis
to evaluate how individuals visually engage with specific regions within a given environment~\cite{rim2021introducing, privitera2000algorithms, wooding2002eye}. 
Metrics such as Time to First Fixation (TFF), fixation counts, and dwell time 
have been effective in assessing attention~\cite{mercier2024quantifying}, 
perception~\cite{tanoubi2021comparing}, and decision-making~\cite{van2015you, rosner2022ambivalence}.
TFF, in particular, has proved important in medical settings for understanding decision-making and situational awareness. For instance, experienced emergency medicine residents in a simulated environment had a shorter TFF for the ``ECG monitor'' AOI (22 vs. 30 seconds) and focused more quickly on critical equipment like the pacing unit, improving emergency decision-making~\cite{tanoubi2021comparing}.
In neonatal care, a study of visual attention during positive pressure ventilation showed that the exhaled tidal volume waveform received the highest total gaze duration and visit count,
compared to other respiratory function monitor parameters, indicating its perceived importance during the procedure~\cite{katz2019visual} .

In a simulated echocardiography study~\cite{hafner2023eye}, 
experts fixated earlier and spent longer dwell times on key AOIs, 
completing ultrasound exams faster than non-experts. 
Research on fourth-year medical students' non-technical skills in emergency care simulations~\cite{anton2023utilizing} revealed that prolonged visual attention on the patient correlated negatively with leadership and communication, whereas a focused gaze on specific elements, like intravenous access, was linked to poorer decision-making and situational awareness.

In summary, AOI analysis plays an important role in understanding medical training, but annotating AOI is challenging, especially with eye-tracking glasses. For one, frequent head movements cause rapid shifts in the video stream's view orientation, 
complicating automated AOI annotation.
Additionally, AOIs are environment-specific and cannot be applied across different simulations. 
Due to these limitations, our study does not focus on AOI-related metrics and instead focuses on more general eye and head tracking measures discussed hereafter.

\subsubsection{Fixation-related metrics}

Fixations are among the most widely used metrics in eye-tracking research, as they capture moments when the gaze remains steady on a single point.
Both fixation duration (time spent fixating) and fixation count (number of fixations)  
have been linked to cognitive load~\cite{fu2024modelling, capogna2020changes, Walter2021}. 
\citet{chen2019can} 
found that surgical residents who spent more time fixating on a feedback screen during needle insertion tasks performed better in later sessions.
Similarly, \citet{capogna2020novice}
observed that expert anesthesiologists performing epidural blocks had fewer but more precise fixations, completing procedures more efficiently, compared to novices.
Another study~\cite{capogna2020changes} showed that hands-on training helped novice anesthesia trainees develop improved focus, leading to fewer but longer fixation duration, indicating enhanced precision.

Building on these findings, we 
propose the following hypothesis:

\begin{enumerate}[leftmargin=9pt]
    \renewcommand\labelenumi{} 
    \item[] \textbf{H1:} Trained practitioners develop different 
        fixation counts (\textbf{H1a}) 
        and fixation durations (\textbf{H1b})
        than untrained practitioners.
\end{enumerate}

\subsubsection{Cognition-related metrics}

A metric for analyzing skill development called Task-Evoked Pupillary Response (TEPR) measures pupil dilation (increase in size) as an indicators of cognitive load. 
TERP has been used as a metric in studies involving medical professionals with different experience levels~\cite{Szulewski2015}.
and has also been used for clinical performance assessment~\cite{Mauriz2023}, 
highlighting its potential for user modeling.
Another relevant metric is Eye Blink Rate (EBR), 
which has been shown to be a good proxy of cognitive flexibility~\cite{Jongkees2016}, 
a key factor for problem-solving, creativity, and learning. 
Furthermore, variations in EBR have been shown to provide valuable information
for assessing cognitive abilities~\cite{Paprocki2017}.
Building on these findings, we 
propose the following hypothesis:

\begin{enumerate}[leftmargin=9pt]
    \renewcommand\labelenumi{} 
    \item[] \textbf{H2:} Trained practitioners develop different 
        TEPR (\textbf{H2a})
        and EBR (\textbf{H2b})
        than untrained practitioners.
\end{enumerate}

\subsubsection{Saccade-related metrics}

Saccadic movements are rapid eye shift that allow redirecting focus from one point to another.
During high precision tasks, saccade amplitude, velocity, and acceleration tend to decrease as the brain prioritizes accuracy and control over speed~\cite{bogacz2010neural}. 
\citet{kessler2023eye} explored saccade-related metrics in a simulated neonatal intubation, tracking the visual focus of pediatric and neonatal practitioners.  
Their findings showed that more experienced practitioners demonstrated better visual attention and situational awareness, though training did not significantly enhance performance. 
\citet{ahmadi2024quantifying} monitored ICU nurses using Tobii Pro Glasses 2 and Empatica E4 devices during 12-hour shifts. 
They found that stress increased both gaze entropy and eye fixations, 
but reduced saccade duration and pupil diameter, 
particularly during high-stress periods like initial handoffs.
Building on these findings, we 
propose the following hypothesis:

\begin{enumerate}[leftmargin=9pt]
    \renewcommand\labelenumi{} 
    \item[] \textbf{H3:} Trained practitioners develop different
    saccade amplitude (\textbf{H3a}), 
    saccade velocity (\textbf{H3b}), 
    and saccade acceleration (\textbf{H3c})
    than untrained practitioners.
\end{enumerate}

\subsection{Head tracking}

Head movements are commonly characterized by acceleration.
\citet{Viriyasiripong2016} measured head movements during simulated laparoscopic suturing surgery 
and found that novices exhibited significantly higher acceleration 
than experts along both vertical and horizontal axes, 
which proved to be a useful metric to evaluate skill development.
Another approach to describe head movement is  angular velocity~\cite{zobeiri2021loss}, 
which reflects the speed of head rotation. 
Additionally, cumulative rotation (the total amount of head movement during a task) 
may provide useful insights for user modeling~\cite{zhao2021atypical}.
Building on these findings, we 
propose the following hypothesis:

\begin{enumerate}[leftmargin=9pt]
    \renewcommand\labelenumi{} 
    \item[] \textbf{H4:} Trained practitioners develop different
    angular velocity (\textbf{H4a}) 
    and cumulative rotation (\textbf{H4b}) 
    than untrained practitioners.
\end{enumerate}

\section{Method}
\label{sec:methodology}

Our aim was to assess skill acquisition during simulation training for breech deliveries,
a childbirth scenario where a baby is born bottom-first instead of head-first. 
This training is part of the Training in Obstetric Emergencies (TUPS) program
which is organized up to four times a year at
Medical Simulation Centre at the University Medical Centre Ljubljana.

TUPS is aimed at specialists in gynecology, obstetrics, and anesthesiology, as well as qualified midwives and nurse anesthetists.
The main goal is to teach professional skills for managing obstetric emergencies, 
emphasizing adherence to professional guidelines and technical executions of standardized procedures. 
The training lasts ten hours and is divided in thematic modules, 
where participants take part in various simulated scenarios. 
The breach delivery training follows the Simulation Education Model in Obstetrics--Pelvic Insertion Parturition (SIP-MV), 
introduced by Steblovnik et al.~\cite{lili2021, lili2014}, which promotes
active participation in a breach delivery scenario in a simulated delivery room. 

\subsection{Participants}
\label{subsec:participants}

We recruited 24 practitioners 
from the TUPS training program.
All participants already finished a 6-year medical degree, 
had prior experience of working in an obstetric room, 
and are currently at various stages of their 5-year specialization training program 
for gynecology ($M=2.72$, $\textit{SD} = 1.32$). 
All participants had normal or corrected-to-normal vision. 

\subsection{Apparatus}
\label{subsec:setup}

\begin{figure*}[!ht]
\centering
\includegraphics[width=\linewidth]{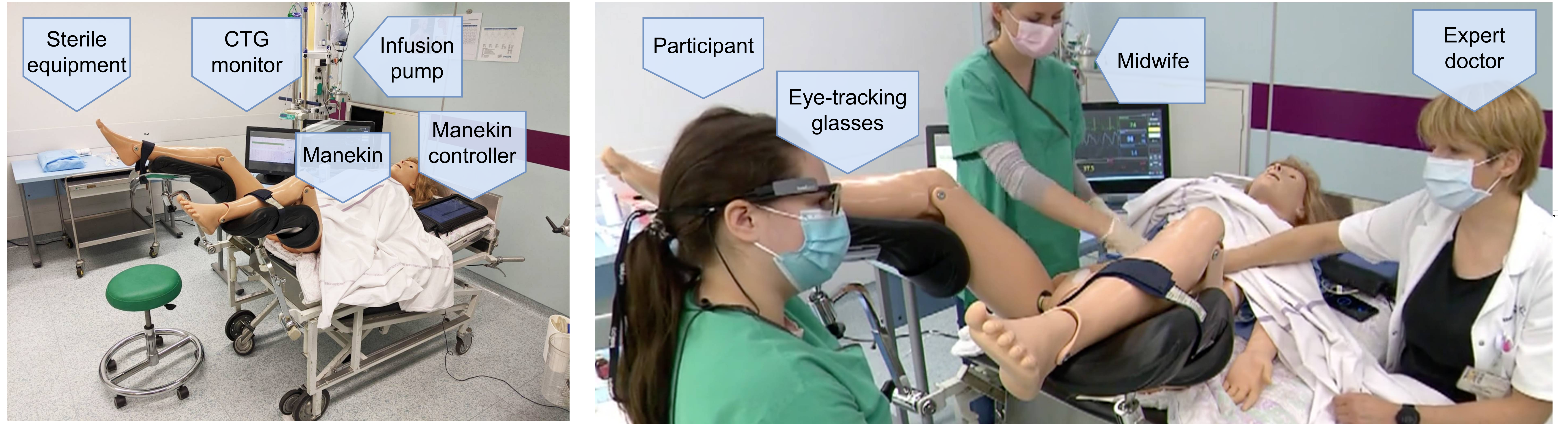}
\caption{Apparatus and delivery room setup. Left: Setup with manikin, controller, CTG monitor, infusion pump, and sterile equipment on a trolley. Right: Expert doctor providing feedback to participant after simulation training.}
\label{fig:apparatus}
\end{figure*}

The breech delivery training was conducted in a simulated delivery room (\autoref{fig:apparatus}) 
which comprised a high-tech NOELLE® S550 manikin that breathes, delivers, speaks, and changes clinical parameters 
under guidance.
The manikin was operated by an expert medical doctor, 
who was also an actor playing the role of a patient, voicing their concerns and hardships.
The room was equipped with a CTG monitor, an infusion pump, a rotating chair on wheels, a trolley, 
and sterile equipment that is usually required for the procedure. 
Besides the expert doctor and the participant, 
a midwife was also present during simulation training (see \autoref{fig:apparatus} right). 
The participant was wearing Tobii Pro Glasses 2, equipped with a Full HD scene camera.\footnote{\url{https://go.tobii.com/Glasses2UM}}
Training sessions were also recorded by two cameras positioned on the ceiling, 
recording a 360 panoramic view of the room and a view overlooking the patient. 

\subsection{Procedure}
\label{subsec:procedure}

The procedure began with a welcome and briefing on the training plan.
Participants were then asked to sign a consent form and fill in a demographic questionnaire collecting information about specialization status, 
prior simulation training experience, and experience with breech delivery. 
Next, participants attended a lecture covering various aspects of breech delivery, 
providing theoretical knowledge and context before proceeding on to the simulation training.

Each participant completed two breech delivery simulation training sessions, 
both conducted in the same delivery room, with each session lasting about seven minutes. 
All of them, including the expert, wore masks as a Covid-19 prevention method as well as to preserve their anonymity during the recordings.
Between the sessions, 
participants engaged in other TUPS-related activities 
that lasted approximately 60 minutes.
Following each session, participants received approximately five minutes of feedback from an expert doctor, focusing on reflective learning.

Before each training session, the eye-tracking glasses were calibrated
using the manufacturer's calibrating software.
If a participant wore corrective glasses, we replaced them with specialized lenses 
that fit directly on the eye-tracking glasses. 

\subsection{Collected data}
\label{subsec:data}

The eye-tracking glasses recorded data at 100\,Hz, 
including pupil size, fixations, saccades, and blinks. 
They also captured head movements at 100\,Hz using a built-in gyroscope and accelerometer,
along with a video stream that enabled visualization of fixations throughout the training sessions.
All recorded data had synchronized timestamps.

\subsection{Task description}
\label{subsec:task}

During the training sessions, participants took the role of a doctor overseeing a natural breech delivery that gets complicated. 
Throughout the procedure, the trainees must demonstrate a set of 14 skills:

\begin{itemize}[leftmargin=13pt]
    \item Introduce themselves to the patient (1).
    \item Gather patient's medical history (2).
    \item Vaginal examination (3).
    \item Prepare for delivery by explaining the procedure to the patient 
        and inserting an intravenous line (4), 
        cleaning the vaginal area (5), 
        and placing a catheter (6).
    \item Determine the appropriate timing and dosage of Sintocinon 
        to address the obstetric arrest (7).
    \item Administer analgesia (8) and perform an episiotomy if needed (9).
    \item Call in additional team members at the appropriate moment to assist with the labor (10).
    \item During labor, apply gentle pressure to the baby’s buttocks 
        to facilitate delivery (11). 
        Identify the active delivery stage when baby’s scapulas become visible (12), 
        at which point the doctor must free the baby's hands (13) 
        to enable the Bracht maneuver assisting birth, without pulling on the baby (14).
\end{itemize}

\begin{figure}[!ht]
\centering
\includegraphics[width=1\linewidth]{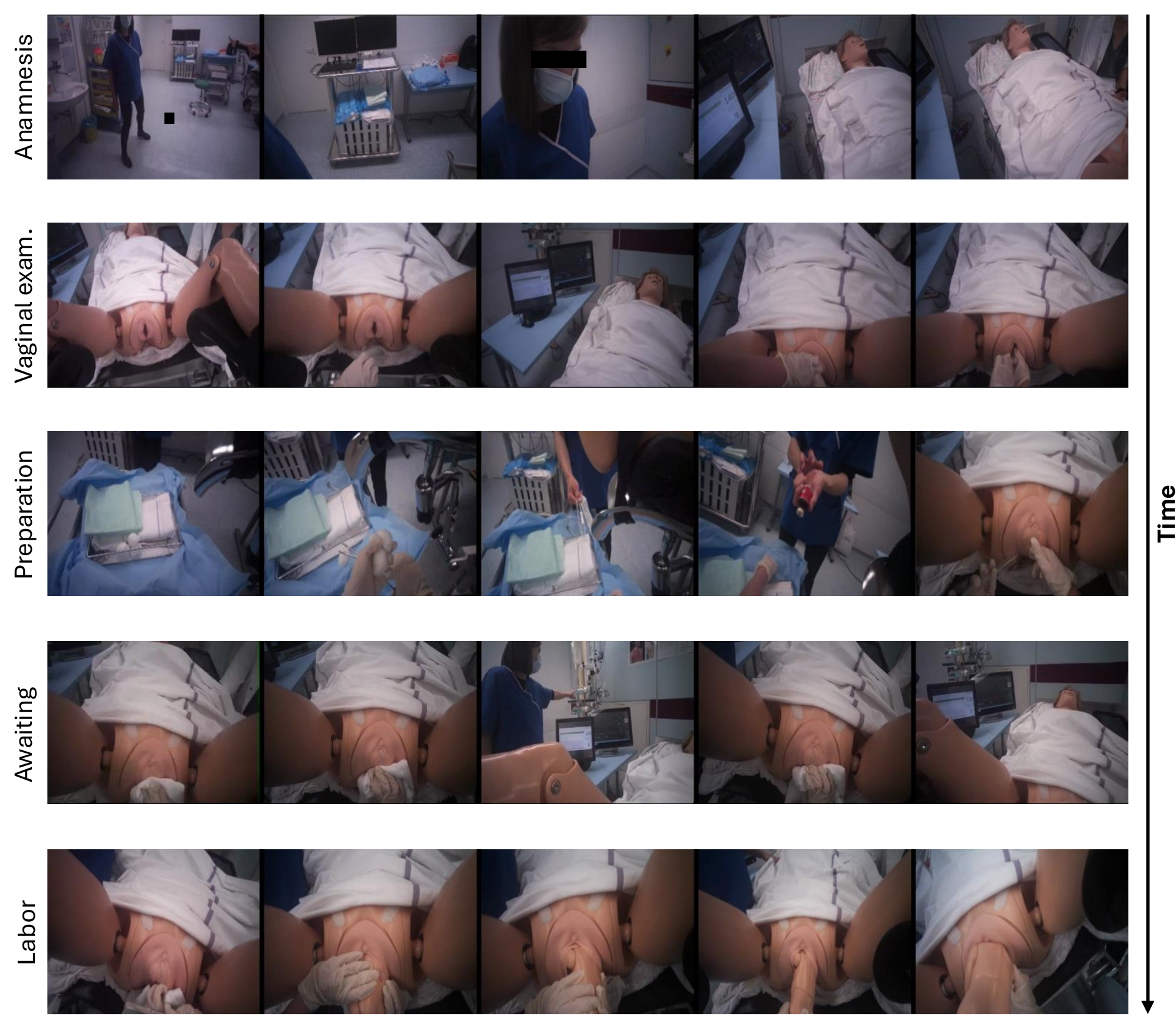}
\caption{
    Example screenshots of a recorded video from different time segments in breech delivery,
    ordered chronologically from top to bottom.
}
\label{fig:intervals-frame-examples}
\end{figure}

\subsection{Analysis}
\label{sec:analysis}

Our goal was to assess the predictive power of eye and head tracking data in assessing medical skills during simulation training.
To achieve this, we first needed to establish a benchmark with the help of an expert medical doctor. 
The doctor viewed all recordings of the training sessions, in no particular order,
and graded each participant on a 1--5 scale for each of the 14 skills (see \autoref{subsec:task}). 

For data analysis, we needed a meaningful way to segment the training sessions 
into smaller time segments to assess changes in the metrics defined in \nameref{sec:relatedwork}.
One approach is to create a new segment for each skill demonstrated. 
However, as some skills are demonstrated sequentially and others in parallel,  
sometimes within a very short time frame, 
we decided to group them under the guidance of the expert doctor.
This process led to the identification of 5 key segments of breech delivery simulation training (see \autoref{fig:intervals-frame-examples}), namely:

\begin{enumerate}[leftmargin=13pt]
    \item \textit{Anamnesis}: Checking patient’s medical history (Skills: 1, 2).
    \item \textit{Vaginal examination}: Assessing the readiness of the cervix and the baby’s position (Skill: 3).
    \item \textit{Preparation}: Setting up the delivery area, 
        ensuring all necessary medical equipment is sterilized 
        and the patient is ready for labor (Skills: 5, 6, 8, 9, 11).
    \item \textit{Awaiting}: The time before going into labor, 
        when instructions are given, 
        intravenous line is inserted, 
        analgesia is admitted, 
        and episiotomy is done (Skills: 4, 8, 10).
    \item \textit{Labor}: The actual process of delivering the baby (Skills: 7, 12, 13, 14)
\end{enumerate}

\section{Results}

\begin{figure}[!ht]
\centering
\includegraphics[width=0.8\linewidth]{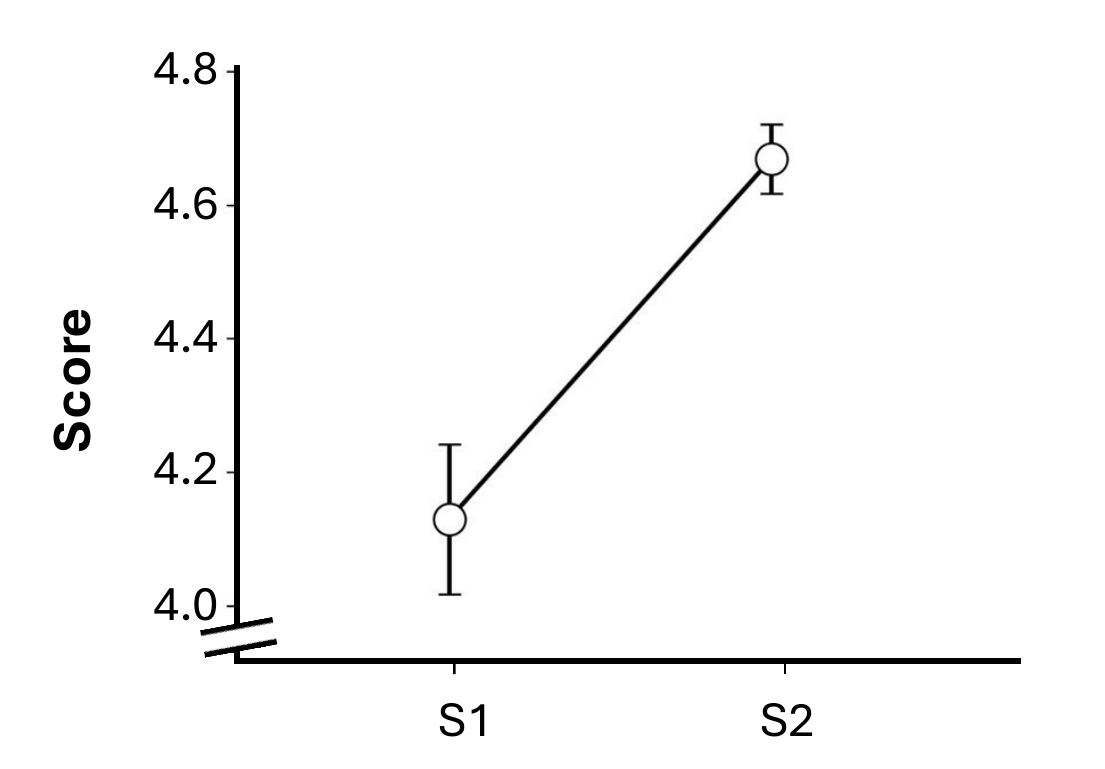}
\caption{Skill scores between sessions (S1: first session, S2: second). Error bars denote standard error of the mean.}
\label{fig:skill-scores-before-after}
\end{figure}

We first checked whether training led to any improvements in skill acquisition. 
We compared the scores assigned by the expert medical doctor
to all participants in both sessions (\autoref{fig:skill-scores-before-after}).
Participants performed better in the second session ($M=4.67$, $SD=0.25$, $Mdn=4.73$) 
compared to the first session ($M=4.13$, $SD=0.55$, $Mdn=4.2$).
A paired $t$-test (two-tails) revealed statistically significant differences:
$t(23) = -5.08, p < .001$.

Next, we compared time spent by participants for each task.
The results are summarized in \autoref{table:interval-duration}.
For Anamnesis, Vaginal examination, and Preparation segments, 
participants performed faster during the second session.
A paired $t$-test (two-tails) was significant
[Anamnesis: $t(23) = 3.80, p < .001$; 
 Vaginal examination: $t(23) = 3.01, p < .01$; 
 Preparation: $t(23) = 4.84, p < .001$].
No differences were found in the Awaiting segment: $t(23) = 0.39, p = .698$.
Participants took significantly more time during Labor in the second session: $t(23) = -5.01, \, p < .001$.

\begin{table}[!ht]
    \caption{
        Duration (in seconds) of each segment 
        for both training sessions (S1: first session, S2: second session).
    }
    \centering
    \resizebox{\linewidth}{!}{
    \begin{tabular}{l c c c c c c}
        \toprule
        \multirow{2}{*}{\textbf{Segment}} & \multicolumn{2}{c}{\textbf{Mean}} & 
        \multicolumn{2}{c}{\textbf{Mdn}} &
        \multicolumn{2}{c}{\textbf{SD}}\\
        \cmidrule(lr){2-3}
        \cmidrule(lr){4-5}
        \cmidrule(lr){6-7}
        & \textbf{S1} & \textbf{S2} & \textbf{S1} & \textbf{S2} & \textbf{S1} & \textbf{S2}\\
        \midrule
        Anamnesis & 56.79 & 41.67 & 54.5 & 42  & 14.14 & 12.24\\
        Vaginal exam. & 44.08 & 31.08 & 40.5 & 31.5 & 15.28 & 13.72\\
        Preparation & 81.12 & 58.04 & 79.5 & 56.5  & 19.11 & 12.48\\
        Awaiting & 105.5 & 101.12 & 108.5 & 106.5 & 33.29 & 39.79\\
        Labor & 39.04 & 67.92 & 30 & 64.5 & 16.50 & 17.89\\
        \bottomrule
    \end{tabular}
    }
    \label{table:interval-duration}
\end{table}

\subsection{Fixation-related metrics}

To ensure the same signal length per segment per participant and to facilitate within-segment comparisons,
we normalized the time in the range of 0 (start of segment) to 1 (end of segment).

\begin{figure*}[!ht]
\centering
\includegraphics[width=1\linewidth]{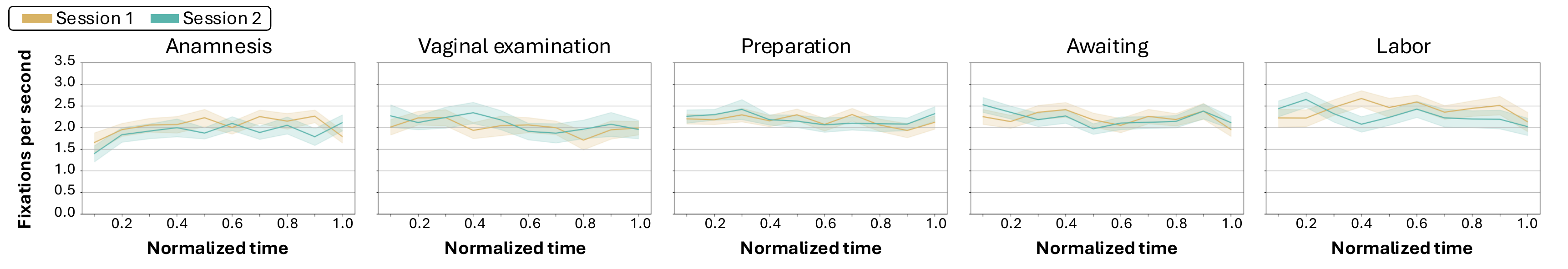}
\caption{Fixation count across different segments. Shaded areas represent the standard error of the mean.}
\label{fig:fixation-count-within-intervals}
\end{figure*}

\begin{figure*}[!ht]
\centering
\includegraphics[width=\linewidth]{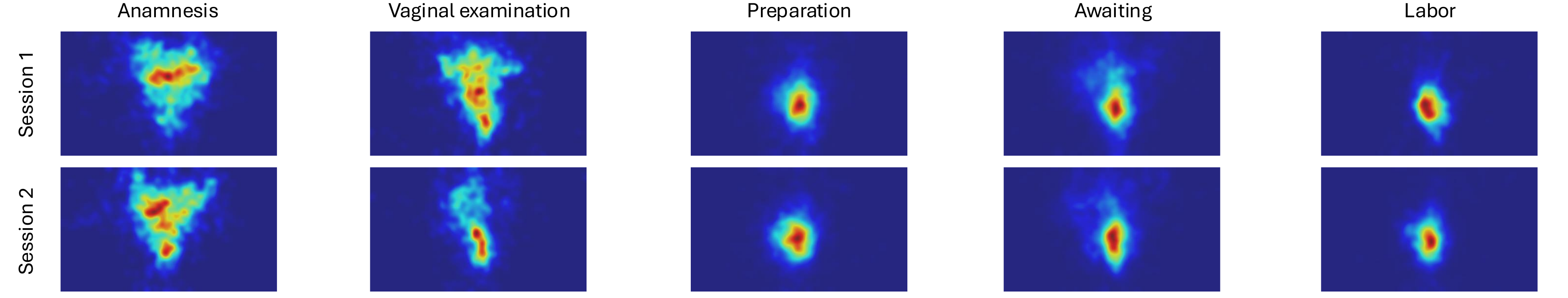}
\caption{Heatmaps of eye fixations.}
\label{fig:fixation-heatmaps}
\end{figure*}

\paragraph{Fixation count.}
Although there were more fixations in the first session,
the paired $t$-test (two-tails) revealed no significant differences 
within any of the segments ($p > .05$).
\autoref{fig:fixation-count-within-intervals} summarizes the results, 
while \autoref{fig:fixation-heatmaps} presents heatmaps of fixation points 
aggregated from all users across time segments.
We can observe more concentrated fixations during the Vaginal examination
in the second session compared to the first.

\begin{figure*}[!ht]
\centering
\includegraphics[width=1\linewidth]{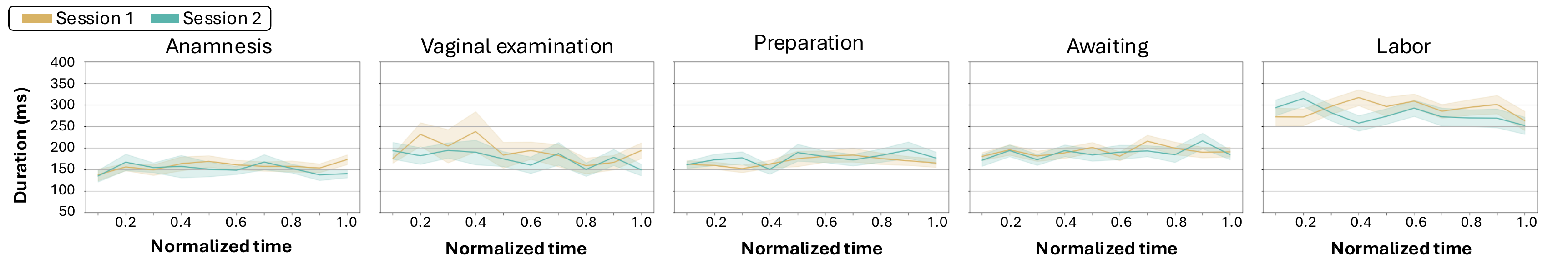}
\caption{Fixation duration across different segments. Shaded areas represent the standard error of the mean.}
\label{fig:fixation-duration-within-intervals}
\end{figure*}

\paragraph{Fixation duration.}
Although fixations lasted longer in the first session,
the paired $t$-test (two-tails) revealed no significant differences 
within any of the segments ($p > .05$).
\autoref{fig:fixation-duration-within-intervals} summarizes the results.

\subsection{Cognition-related metrics}

\begin{figure*}[!ht]
\centering
\small
\includegraphics[width=1\linewidth]{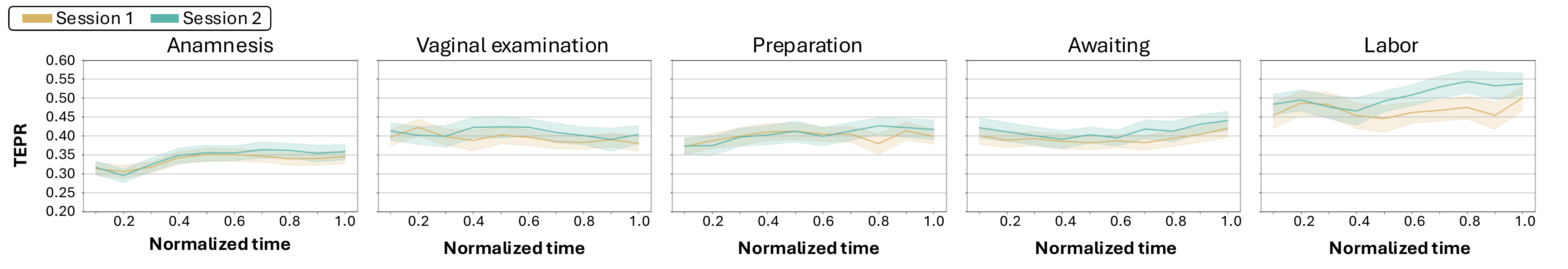}
\caption{Pupil size across different segments. Shaded areas represent the standard error of the mean.}
\label{fig:pupils-within-intervals}
\end{figure*}

\paragraph{Task-Evoked Pupillary Response.}
We took the average of the left and right pupil sizes, 
and normalized them using Min-Max normalization, 
considering the entire signal from both sessions. 
While TEPR was higher in the second session,
the paired $t$-test (two-tails) was significant only during Labor: $t(23) = -2.41, p = .017$.
\autoref{fig:pupils-within-intervals} summarizes the results.
 
\begin{figure*}[!ht]
\centering
\includegraphics[width=1\linewidth]{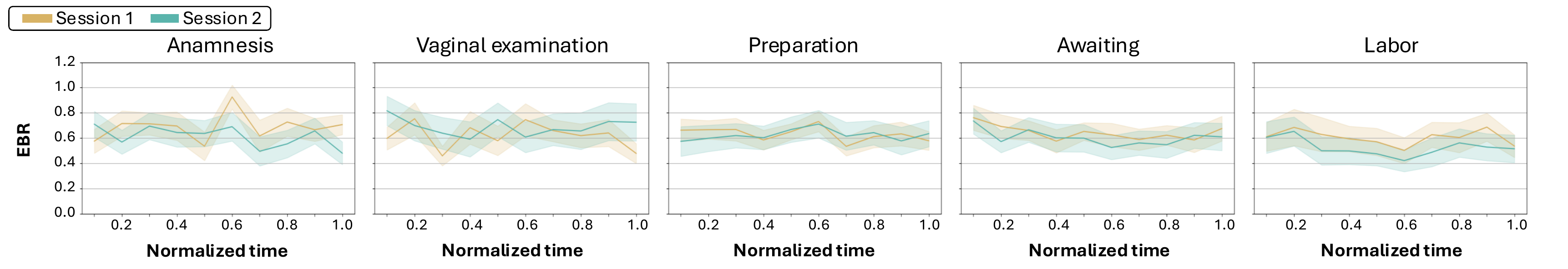}
\caption{Blink rate across different segments. Shaded areas represent the standard error of the mean.}
\label{fig:blinks-within-intervals}
\end{figure*}

\paragraph{Eye Blink Rate.}
We excluded any blinks with a duration of less than 100\,ms
as this threshold is considered the minimum duration for a valid blink~\cite{huette2016blink, frank2015validating}.
While EBR was higher in the first session, especially during Anamnesis and Labor, 
no significant differences were observed within any of the segments ($p > .05$).
\autoref{fig:blinks-within-intervals} summarizes the results.

\subsection{Saccade-related metrics}

\begin{figure*}[!ht]
\centering
\includegraphics[width=\linewidth]{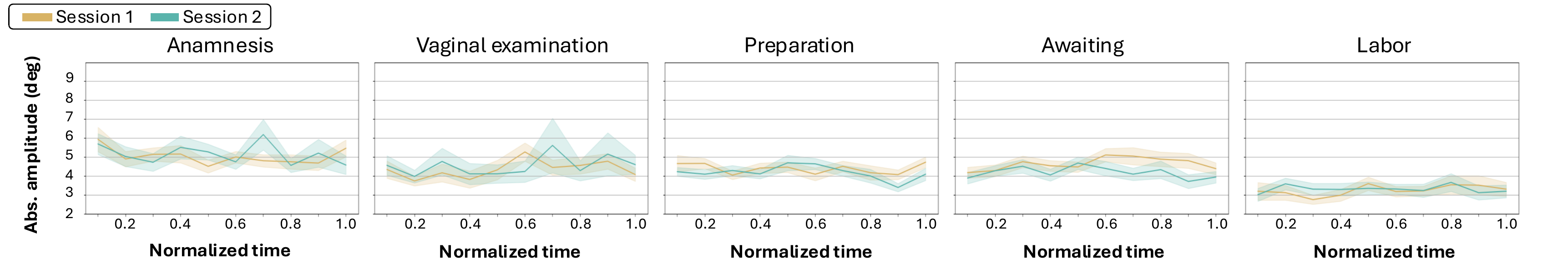}
\caption{Saccade amplitude across different segments. Shaded areas represent the standard error of the mean.}
\label{fig:saccade-amplitude-within-intervals}
\end{figure*}

\paragraph{Saccade amplitude.}
While saccade amplitude was higher for the first session,
the paired $t$-test (two-tails) was significant 
only during Awaiting: $t(23) = 2.9465, p < .01$.
\autoref{fig:saccade-amplitude-within-intervals} summarizes the results.

\begin{figure*}[!ht]
\centering
\includegraphics[width=1\linewidth]{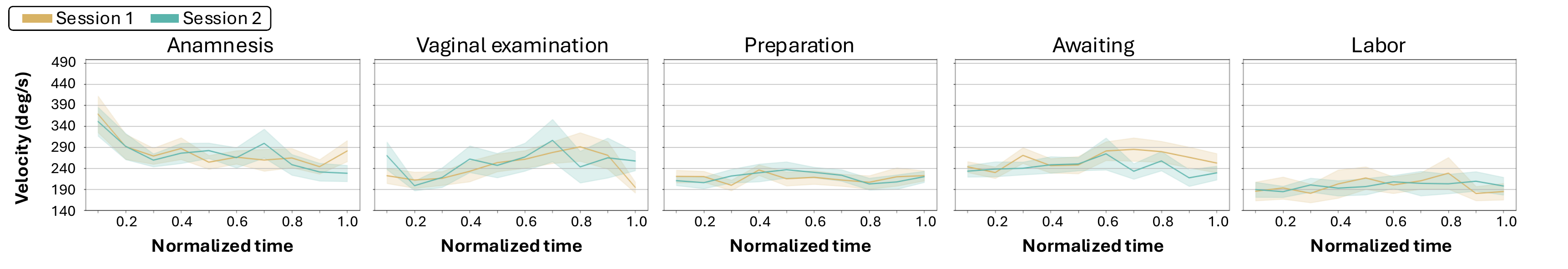}
\caption{Saccade velocity across different segments. Shaded areas represent the standard error of the mean.}
\label{fig:saccade-velocity-within-intervals}
\end{figure*}

\paragraph{Saccade velocity.}
While saccade velocity was faster in the first session,
the paired $t$-test (two-tails) revealed no significant differences 
within any of the segments ($p > .05$).
\autoref{fig:saccade-velocity-within-intervals} summarizes the results.

\begin{figure*}[!ht]
\centering
\includegraphics[width=1\linewidth]{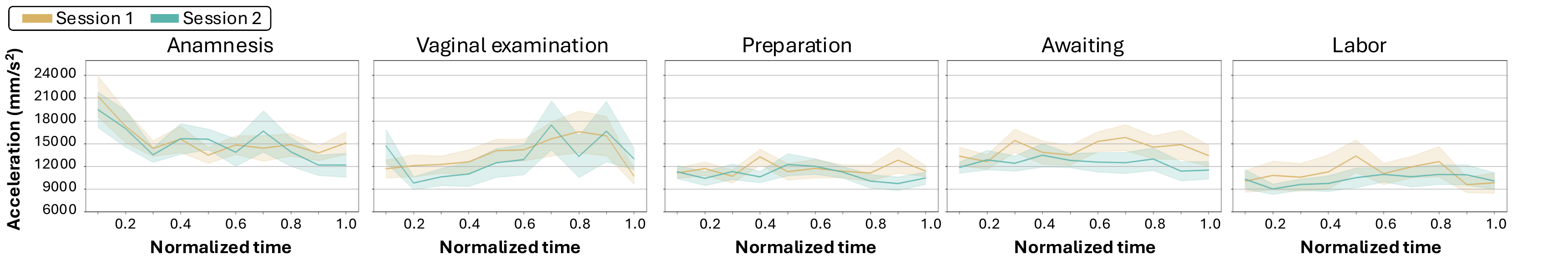}
\caption{Saccade acceleration across different segments. Shaded areas represent the standard error of the mean.}
\label{fig:saccade-acceleration-within-intervals}
\end{figure*}

\paragraph{Saccade acceleration.}
While saccade acceleration was higher for the first session,
the paired $t$-test (two-tails) was significant 
only during Awaiting: $t(23) = 2.78, p < .01$.
\autoref{fig:saccade-acceleration-within-intervals} summarizes the results.

\subsection{Head-related metrics}

\begin{figure*}[!ht]
\centering
\includegraphics[width=1\linewidth]{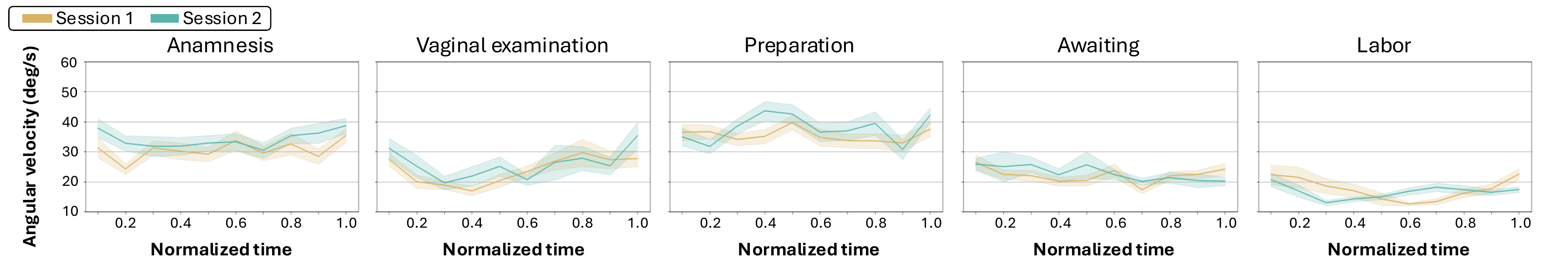}
\caption{Angular velocity of head movements. Shaded areas represent the standard error of the mean.}
\label{fig:head-angular-velocity-within-intervals}
\end{figure*}

\paragraph{Angular velocity.}
While angular velocity was higher for the first session,
the paired $t$-test (two-tails) was significant 
only during Anamnesis: $t(23) = -2.897, p < .01$.
\autoref{fig:head-angular-velocity-within-intervals} summarizes the results.

\paragraph{Cumulative rotation}

\begin{figure*}[!ht]
\centering
\includegraphics[width=1\linewidth]{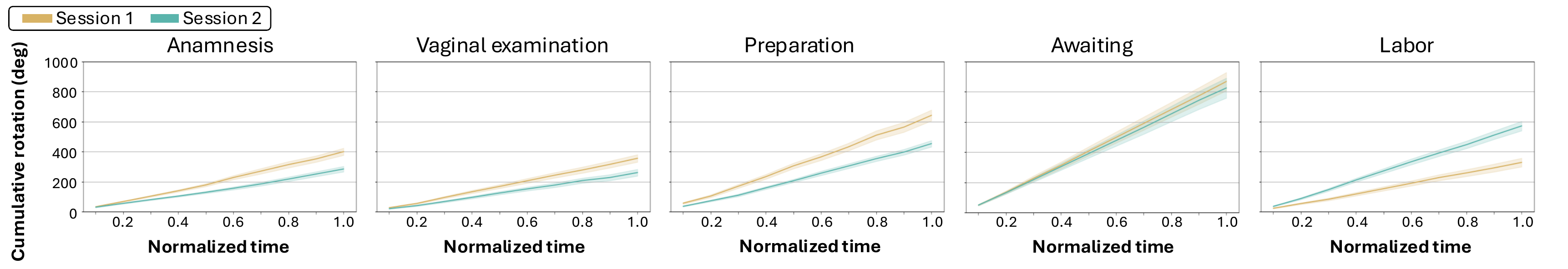}
\caption{Cumulative rotation of head movements. Shaded areas represent the standard error of the mean.}
\label{fig:head-cumulative-rotation-within-intervals}
\end{figure*}

Cumulative rotation was significantly higher for the first session,
as corroborated by the paired $t$-test (two-tails),
during Anamnesis ($t(23) = 10.30, p < .001$),
Vaginal examination ($t(23) = 7.40, p < .001$),
and Preparation ($t(23) = 10.94, p < .001$).
No statistically significant differences were found during Awaiting ($p > .05$).
Finally, cumulative rotation was significantly lower during Labor ($t(23) = -12.08, p < .001$). 
\autoref{fig:head-cumulative-rotation-within-intervals} summarizes the results.

\section{Machine Learning models}
\label{sec:data-representations-and-models}

We observed significant improvements in skill acquisition but most eye/head related metrics revealed no statistically significant differences between training sessions. 
Therefore, aimed at further investigating the role of eye and head movements for user modeling, 
we trained Machine Learning classifiers to tell trained and untrained practitioners apart (binary classification task).

We trained Support Vector Machine (SVM) classifiers,
since they have been widely used for eye-tracking classification tasks~\cite{lim2022eye, haddad2024good}
given their efficiency and adequacy in handling small sample sizes.
We employed AutoML with Bayesian Optimization to tune the following SVM hyperparameters: 
kernel type\footnote{All polynomial kernels had a degree of 3.} 
$\in$ \{Linear, RBF, Polynomial\}, 
regularization parameter $C$ within $0.1 \leq C \leq 100$, 
and decay for non-linear kernels within $0.01 \leq \gamma \leq 10$.

We divided each segment into smaller non-overlapping sampling windows.
For each window, we engineered a feature vector specific to each modality 
(\autoref{table:handcrafted-features}) consisting of: Min, Max, Mean, Mdn, and SD of the values in each window. 
As in the previous section, time was normalized between 0 and 1, 
relative to the start and end of each segment. 
We also concatenated a numerical code to the feature vectors, 
to inform the model about the segment from which each sample originates: 
\{1:\,Anamnesis, 2:\,Vaginal~examination, 3:\,Preparation, 4:\,Awaiting, 5:\,Labor\}. 

The target label, to predict as model output, was
either `0', representing the first session (untrained practitioners), 
or `1', representing the second session (trained practitioners). 
We first trained each model on individual segments, 
to assess how effectively each of them contributes to discriminating between practitioners. 
We then considered the concatenation of all segments at once.
In any case, we used 80\% of the data as the training set, 
10\% for validation, and 10\% as the test set.
Note that each modality has a different number of feature vectors 
(\autoref{table:handcrafted-features}) since the sampling windows are different. 

\begin{table}[!ht]
    \caption{List of handcrafted features for SVM classifiers.}
    \centering
    \small
        \begin{tabular}{lll}
            \toprule
            \textbf{Modality} & \textbf{Features} & \textbf{Sampling window} \\
            \midrule
            \multirow{4}{*}{Pupil} 
            & Timestamp & \multirow{4}{*}{\makecell{100 data points (1\,s) \\ 10751 feat vectors}} \\
            & X and Y coordinates & \\
            & Normalized pupil size & \\
            & Segment code & \\
            \hline
            \multirow{4}{*}{Fixation} 
            & Timestamp & \multirow{4}{*}{\makecell{10 successive fixations \\ 3361 feat vectors}} \\
            & X and Y coordinates & \\
            & Duration & \\
            & Segment code & \\
            \hline
            \multirow{5}{*}{Saccade} 
            & Timestamp & \multirow{5}{*}{\makecell{10 successive saccades \\ 8729 feat vectors}} \\
            & Amplitude & \\
            & Peak velocity & \\
            & Peak acceleration & \\
            & Segment code & \\
            \hline
            \multirow{3}{*}{Blinks} 
            & Timestamp & \multirow{3}{*}{\makecell{10 successive blinks \\ 1051 feat vectors}} \\
            & Duration & \\
            & Segment code & \\
            \hline
            \multirow{3}{*}{Head} 
            & Timestamp & \multirow{3}{*}{\makecell{100 data points (1\,s) \\ 15107 feat vectors}} \\
            & Rotational speed in X, Y, Z & \\
            & Segment code & \\
            \bottomrule
        \end{tabular}
    \label{table:handcrafted-features}
\end{table}

\subsection{Classification Performance}
\label{sec:classification-performance}

\autoref{table:classification-performance} reports the weighted $F_1$ and AUC scores of each classifier, highlighting the discriminating power of each modality in distinguishing between trained and untrained practitioners. Overall, the head-based classifiers performed best. The highest performance was observed for the Labor segment, with 85\% $F_1$ and 86\% AUC, followed by 77\% $F_1$ and 84\% AUC for the Preparation segment. Among the eye-based classifiers, the highest performance was also observed for the Labor segment, with an $F_1$ of 77\% and AUC of 85\% using pupillary responses and 67\% $F_1$ and 80\% AUC using fixation data. Blink data also performed similarly. These results are particularly encouraging, as discussed in the next section.

\begin{table}[!ht]
    \caption{
        Classification performance results.
        Best result in bold. 
        Second best result underlined.
    }
    \centering
    \resizebox{\linewidth}{!}{
    \begin{tabular}{l l c c c c c}
        \toprule
        \multirow{2}{*}{\textbf{Modality}} & \multirow{2}{*}{\textbf{Segment}} & \multicolumn{3}{c}{\textbf{Hyperparameters}} & \multirow{2}{*}{\textbf{Adj. F$_1$}} & \multirow{2}{*}{\textbf{AUC}}\\
        \cmidrule(lr){3-5}
        & & \textbf{kernel} & \bm{$C$} & \bm{$\gamma$}\\
        \midrule
        \multirow{5}{*}{Pupil} & Anamnesis & linear & 47.35 & 0.01 & 0.68 & 0.74\\
        & Vaginal exam. & linear & 12.58 & 0.01 & 0.58 & 0.69\\
        & Preparation & linear & 100.0 & 10.0 & \ul{0.70} & \ul{0.79}\\
        & Awaiting & RBF & 25.21 & 0.58 & 0.67 & 0.74\\
        & Labor & linear & 100.0 & 0.85 & \textbf{0.77} & \textbf{0.85}\\
        \cmidrule{2-7}
        & All segments & RBF & 19.46 & 0.40 & 0.65 & 0.70\\
        \hline
        \multirow{5}{*}{Fixation} & Anamnesis & linear & 6.64 & 0.01 & 0.62 & 0.64\\
        & Vaginal exam. & RBF & 1.70 & 6.29 & 0.44 & 0.54\\
        & Preparation & linear & 54.39 & 0.01 & \ul{0.62} & \ul{0.81}\\
        & Awaiting & RBF & 32.98 & 0.01 & 0.59 & 0.61\\
        & Labor & linear & 73.52 & 0.80 & \textbf{0.67} & \textbf{0.80}\\
        \cmidrule{2-7}
        & All segments & RBF & 13.50 & 0.01 & 0.56 & 0.58\\
        \hline
        \multirow{5}{*}{Saccade} & Anamnesis & linear & 44.23 & 10.0 & 0.59 & 0.63\\
        & Vaginal exam. & linear & 27.37 & 0.62 & 0.61 & \ul{0.64}\\
        & Preparation & linear & 56.50 & 0.01 & \ul{0.63} & \textbf{0.69}\\
        & Awaiting & RBF & 100.0 & 0.02 & 0.54 & 0.56\\
        & Labor & linear & 27.36 & 0.62 & \textbf{0.64} & \textbf{0.69}\\
        \cmidrule{2-7}
        & All segments & RBF & 93.19 & 0.01 & 0.55 & 0.57\\
        \hline
        \multirow{5}{*}{Blink} & Anamnesis & RBF & 49.38 & 0.32 & 0.42 & 0.52\\
        & Vaginal exam. & RBF & 0.64 & 9.89 & 0.42 & 0.44\\
        & Preparation & RBF & 100.0 & 7.33 & \ul{0.52} & 0.48\\
        & Awaiting & polynomial & 21.58 & 2.83 & 0.39 & \ul{0.58}\\
        & Labor & RBF & 15.92 & 0.03 & \textbf{0.66} & \textbf{0.84}\\
        \cmidrule{2-7}
        & All segments & RBF & 27.36 & 0.62 & 0.46 & 0.52\\
        \hline
        \multirow{5}{*}{Head} & Anamnesis & linear & 77.60 & 9.97 & 0.63 & 0.69\\
        & Vaginal exam. & linear & 81.10 & 0.01 & 0.65 & 0.72\\
        & Preparation & linear & 100.0 & 0.56 & \ul{0.77} & \ul{0.84}\\
        & Awaiting & RBF & 0.1 & 0.49 & 0.47 & 0.48\\
        & Labor & linear & 31.76 & 10.0 & \textbf{0.85} & \textbf{0.86}\\
        \cmidrule{2-7}
        & All segments & RBF & 100.0 & 0.15 & 0.55 & 0.58\\
        \bottomrule
    \end{tabular}
    }
    \label{table:classification-performance}
\end{table}

\section{Discussion}
\label{sec:results}

Participants  demonstrated improved expertise after both training sessions, as reflected in their skill assessment scores.
While eye and head movements effectively distinguish between trained and untrained practitioners, our findings suggest that certain segments of the training process—particularly Labor—play a more critical role in skill assessment. Furthermore, eye and head movement data provide valuable insights into different aspects of skill development.

During the second session, we observed an increase in TEPR during Labor, indicating both higher cognitive load and engagement~\cite{van2016pupil}. This segment also exhibited longer fixation durations but fewer fixation counts, suggesting a heightened level of concentration as participants became more skilled~\cite{aljehane2023studying, mahanama2022eye}. We attribute this behavior to trained practitioners being able to consider more options before making a critical decision, as indicated by greater pupil dilation during the second session~\autoref{fig:pupils-within-intervals}. This also explains the longer duration of Labor after the first session~\autoref{table:interval-duration} while trained participants performed faster in the rest of the segments.
Consequently, we reject \textbf{H1a} and \textbf{H1b}, concluding that trained practitioners exhibit similar fixation counts and fixation durations as untrained practitioners. We partially accept \textbf{H2a}, as TEPR showed significant differences during Labor, but we reject \textbf{H2b} due to the lack of differences in EBR.

Saccadic movements primarily revealed differences between the two training sessions 
during Awaiting and Labor. 
Lower saccade amplitude and increased velocity acceleration in the first session suggests that participants developed quicker yet more focused movements 
as their skills improved after the first training session.
As a result, we reject \textbf{H3b}, as trained and untrained practitioners demonstrated similar saccade velocity. However, we partially accept \textbf{H3a} and \textbf{H3c}, as saccade amplitude and acceleration were significantly different during Awaiting.

Finally, our findings confirm that head movements are a reliable indicator of skill progression.
The increase in angular velocity during Anamnesis, Vaginal examination, and Preparation 
suggests greater fluency in executing these preparatory steps before Labor. 
Additionally, the decrease in cumulative rotation during the second session (except for Labor)
indicates increased focus, resulting in reduced head movements to complete the tasks.
Based on this, we partially accept \textbf{H4a} and \textbf{H4b},
since angular velocity was significantly different in Amnesis,
and cumulative rotation was significantly different in all segments except Awaiting.

\section{Limitations and future work}
\label{sec:limitations_future_work}

We acknowledge a relatively small sample size in our study (24 participants),
however this is a common challenge in medical research~\cite{mitani2020small}, 
given the difficulty of recruiting professionals~\cite{bruneau2021recruitment}.
Another limitation of our study is that we normalized the time of our collected signals,
to facilitate the comparisons within each segment, 
which assumes that each practitioner spent similar time in each segment.
While this holds for most cases, 
the Awaiting segment showed greater variability (see \autoref{table:interval-duration}).

We should point out that 
accurately tracking saccadic movements requires an eye tracker operating at least at 200\,Hz. 
These high-frequency eye trackers are usually available in stationary form only.
Due to the physical constraints of the delivery room simulation, 
we used eye-tracking glasses, that operated at 100\,Hz. 
Despite this, our classification results indicate that meaningful data can still be extracted at this frequency. 
Future advances in eye-tracking technology are expected to enhance accuracy further.

Additionally, our findings highlight the potential of SVM models for accurate skill classification.
We should note that, in addition to the SVM classifiers, 
we trained Recurrent Neural Network (RNN) and XGBoost models, 
also utilizing AutoML with Bayesian Optimization, but they did not perform well. 
The RNN model had an input layer of N dimensions (where N is the size of the sampling window, see \autoref{table:handcrafted-features})
followed by a hidden LSTM layer using either hyperbolic tangent or ReLU activation. 
The embedding size of the hidden layer ranged from 50 to 100, in increments of 10. 
This was followed by a dropout layer with values between 0.1 and 0.5, in increments of 0.1. 
Then a fully connected layer with a single neuron and sigmoid activation was added for the final output. 
The candidate learning rates were selected from the set \(\{10^{-n} \mid n = 3, 4\}\). 
For the XGBoost classifiers, the parameter space included the number of estimators (50--500), maximum depth (3--10), learning rate (0.01--0.3, using a log-uniform distribution), subsample ratio (0.5--1.0), and the fraction of features considered per split (0.5--1.0). The main reason for lower performance using LSTM is that the number of samples (time series) per segment is not large enough for RNN model training. One possibility for future work would be to apply data augmentation strategies.
Future work should also consider multimodal fusion, 
combining features from eye and head movements to improve further model performance.
Moreover, given the discriminative power of head movements, 
further analysis of specific types, such as neck extension and lateral bending~\cite{chen2019atomic}, 
could provide deeper insights. 
Additionally, exploring a range of different clinical tasks could provide valuable insights into the robustness of our methodology and results.

\section{Conclusion}
\label{sec:conclusion}

We conducted a user study exploring eye and head tracking 
to understand skill development in clinical settings, specifically
during simulated baby delivery. 
Our results show that eye and head movements can
effectively distinguish trained from untrained practitioners with remarkable performance, with 
some tasks (e.g., Labor) being more informative than others.
These findings lay the groundwork for computational models 
that support implicit and objective skill assessment using commodity eye-tracking glasses, 
complementing traditional evaluation methods
in clinical settings such as explicit and objective questionnaires.
Ultimately, our results pave the way to faster and less biased assessment 
of the skills of medical doctors in training activities.

\begin{acks}
This research is supported by the European Innovation Council Pathfinder program (SYMBIOTIK project, grant 101071147), 
the Slovenian Research Agency (grants N2-0354, BI-NO/25-27-007, P5-0433, IO-0035, J5-50155, J7-50096), 
and the CogniCom (grant 0013103) program of the University of Primorska. 
Authors would like to thank all medical and technical support staff of the Medical Simulation Unit of the University Medical Center Ljubljana.
\end{acks}



\end{document}